# Use of AI Tools: Guidelines to Maintain Academic Integrity in Computing Colleges

*Hatem M. El-boghdadi [1]*
*Toqeer Ali Syed[1, †]*
*Ali Akarma[1]*
*and Qamar Wali [2]*

[1]Faculty of computer and information System,
Islamic University of Madina, Saudi Arabia
[2] Faculty of Artificial Intelligence
Multimedia University, Malaysia

## Abstract

The rapid adoption of AI tools such as ChatGPT has significantly transformed academic practices, offering considerable benefits for both students and faculty in computing disciplines. These tools have been shown to enhance learning efficiency, academic self-efficacy, and confidence. However, their increasing use also raises pressing concerns regarding the preservation of academic integrity -- an essential pillar of the educational process. This paper explores the implications of widespread AI tool usage within computing colleges, with a particular focus on how to align their use with the principles of academic honesty. We begin by classifying common assessment techniques employed in computing education and examine how each may be impacted by AI-assisted tools. Building on this foundation, we propose a set of general guidelines applicable across various assessment formats to help instructors responsibly integrate AI tools into their pedagogy. Furthermore, we provide targeted, assessment-specific recommendations designed to uphold educational objectives while mitigating risks of academic misconduct. These guidelines serve as a practical framework for instructors aiming to balance the pedagogical advantages of AI tools with the imperative of maintaining academic integrity in computing education. Finally, we introduce a formal model that provides a structured mathematical framework for evaluating student assessments in the presence of AI-assisted tools.

## 1. Introduction

Since its launch in 2022, ChatGPT has become the fastest-growing application in internet history [3], amassing nearly 100 million users by January 2023 and attracting approximately 1.8 billion monthly visitors to its website as of recent reports [4,5]. Its widespread adoption has extended deeply into academic environments, where students increasingly rely on ChatGPT for a variety of educational purposes—including understanding complex concepts, solving technical problems, generating ideas, and composing written work.

---

[†] Corresponding Author's Email: toqeer@iu.edu.sa

While platforms like ChatGPT offer substantial advantages -- such as saving time on routine tasks, assisting with idea generation, and supporting conceptual understanding--concerns about academic integrity have emerged. Some academics argue that reliance on such tools could undermine essential educational outcomes and devalue independent student work. In response, certain institutions have taken a restrictive stance, opting to ban the use of ChatGPT entirely [6], while others continue to debate its appropriate role in education [7]. In this context, our aim is to advocate for the responsible and guided use of AI tools in academic settings, especially within computing disciplines. By implementing structured guidelines, educators can preserve academic integrity while embracing the pedagogical benefits of emerging AI technologies.

The main question that arises: Is allowing use of tools such as ChatGPT for students would impact the academic integrity? The answer greatly depends on how AI tools are used. Use of AI tools could impact academic integrity, but it does not have to. With clear guidelines to students on how to use such tools and guidelines to instructors on how to complete different types of assessment, AI tools would become an assisting tool like calculators, students should not miss to use.

However, the misuse of such tools could affect academic integrity negatively. AI generated work could be submitted by students as their own work without understanding or effort. Lack of critical thinking could be the result if students rely heavily on these tools. Even, using the AI tools without permission breaks the academic rules.

The rapid integration of AI tools such as ChatGPT into academic settings has introduced both significant educational benefits and serious concerns regarding academic integrity. While these tools support student learning by improving efficiency, comprehension, and creativity, their unregulated use may lead to misuse, plagiarism, and diminished critical thinking. As illustrated in Figure 1, this dual impact has prompted varying institutional responses, ranging from outright bans to ongoing policy debates. The core problem lies in the absence of structured guidelines that enable the ethical use of AI tools while preserving the integrity and objectives of the education process, particularly in computing disciplines where such tools are most frequently applied.

In this paper, we are interested in helping the academic community to maintain academic integrity in computing colleges. Given that the use of AI by students in computing colleges is a fact and cannot be stopped and considering the wide use of AI tools in computing colleges, we take a step forward to take advantages of AI tools while maintaining the academic integrity. First, we classify the assessment techniques used in computing colleges according to its types. Then, we introduce general guidelines that can be followed with most types of assessment. Finally, we consider each type of assessment and introduce specific guidelines for that assessment as to maintain the target of education process and hence the academic integrity. These guidelines are meant to be used by instructors in computing colleges.

This paper is organized as follows. The next section presents how different factors could affect the use of AI tools in the education process. Section 3 considers different types of assessment in specifically computing colleges. In section 4, we introduce some general guidelines that can be followed by instructors to all assessment types and specific guidelines for each type of assessment. Section 5 introduces a formal model that provides a structured mathematical framework for evaluating student assessments. Finally, in section 6, we make some concluding remarks.

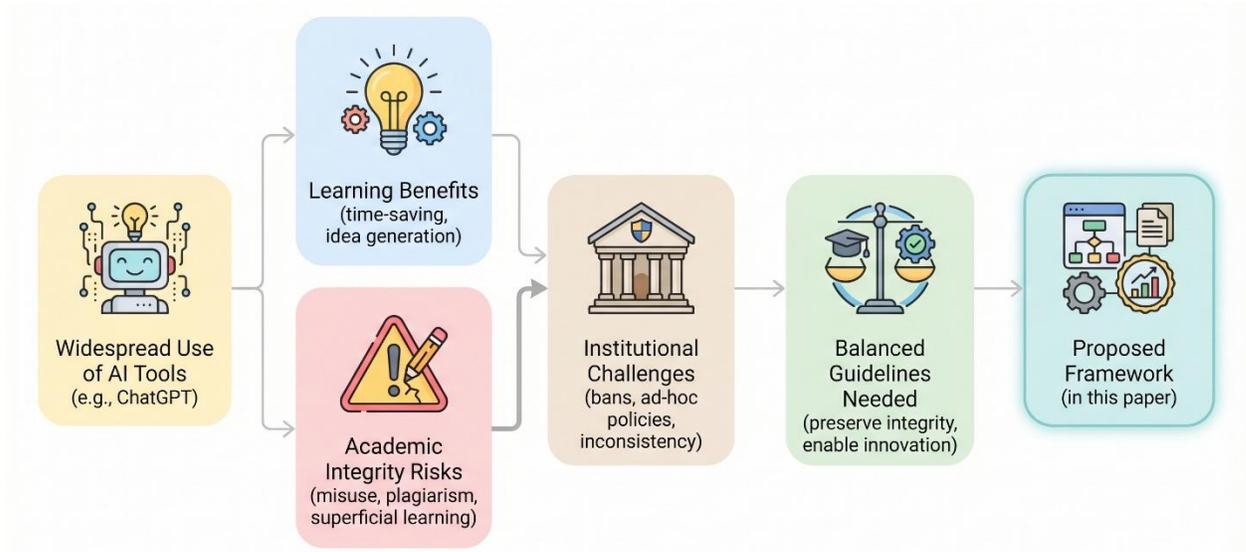

Figure 1: Problem context illustrating the dual impact of AI tools in computing education and motivating the proposed framework.

## 2. Effect of Using AI tools on Education process

One of the primary objectives of the education process is the development of students' cognitive, analytical, and technical skills. Critical thinking is not acquired instantly; rather, it is developed progressively as students solve problems, propose designs, analyze scenarios, and reflect on their own reasoning processes. These activities are central to learning outcomes in computing and engineering education.

If students rely entirely on AI tools to complete their academic work—such as writing essays, preparing literature surveys, solving analytical problems, or automatically generating source code—the educational value of these activities may be significantly diminished. While such reliance may allow students to complete assignments more quickly and potentially achieve higher grades, it can undermine the development of essential skills, including problem-solving ability, conceptual understanding, and independent reasoning. Consequently, unrestricted or undisclosed use of AI tools risks weakening the core objectives of the education process [28].

On the other hand, when AI tools are used responsibly and transparently, they can provide meaningful educational benefits. AI systems can assist students by explaining complex concepts, suggesting alternative solution strategies, supporting idea generation, and improving the clarity and structure of written work. In this sense, AI tools can function as supportive learning aids rather than replacements for student effort, similar to calculators, simulators, or integrated development environments used in technical disciplines.

Recent studies have examined several factors influencing the adoption of AI tools such as ChatGPT in educational contexts. These factors include time-saving benefits [8], academic performance considerations [9], electronic word-of-mouth effects [10,11], peer and social influence [12–14], self-esteem and self-perception [15], academic self-efficacy [16,17], and perceived academic stress [18,19]. Prior research indicates that these factors collectively contribute to increased AI tool usage in academia, particularly among students seeking efficiency and academic support. While such findings highlight the growing prevalence of AI-assisted learning, they also reinforce the need for structured guidelines to ensure that AI usage supports learning objectives rather than undermining academic integrity [20–24].

Recent studies further highlight that the rapid adoption of generative AI tools in higher education presents both pedagogical opportunities and integrity-related challenges [25,26]. Systematic reviews and policy-oriented analyses emphasize that while tools such as ChatGPT can support learning efficiency and conceptual understanding, they also necessitate assessment redesign, transparency requirements, and verification mechanisms to preserve academic integrity [27-30]. Emerging research consistently argues that traditional assessment formats are increasingly vulnerable in the presence of generative AI, reinforcing the need for structured guidelines that balance innovation with ethical and educational objectives [31,32].

## 2.1 Equity, Accessibility, and Bias in AI-Assisted Education

While AI tools offer significant educational advantages, their integration into academic environments raises important concerns related to equity, accessibility, and potential bias. Not all students have equal access to advanced or premium AI tools, such as paid versions of large language models [33]. This disparity can introduce unintended advantages for students with greater financial resources, potentially compromising fairness in assessment outcomes.

To address accessibility concerns, educational institutions should design assessments that do not depend on exclusive features of premium AI tools. Guidelines may encourage the use of freely available AI systems or ensure that AI assistance does not confer a disproportionate advantage to any group of students [34]. Transparency requirements, such as mandatory disclosure of AI usage, can further support equitable evaluation.

In addition, AI-generated content may reflect biases present in training data, which can influence explanations, examples, or suggested solutions. Such biases may affect the quality or neutrality of AI-assisted outputs, particularly in open-ended or evaluative tasks. To mitigate these risks, verification mechanisms and human-in-the-loop assessment practices are essential [35]. By combining AI-assisted learning with instructor oversight and critical evaluation, institutions can reduce bias-related risks while maintaining fairness and academic rigor.

# 3. Assessments in Computing Colleges

In this section, we identify and describe common types of assessments used in computing colleges and discuss how the availability of AI tools like ChatGPT can affect their integrity [36]. These assessments typically fall into two broad categories: in-class (supervised) and out-of-class (unsupervised) assessments. Each type presents unique challenges and opportunities regarding the ethical and effective use of AI-assisted tools [37, 38].

## 3.1 Homework Assignments

Homework assignments are among the most frequent out-of-class assessment types. These may include coding tasks, theoretical questions, algorithm analysis, or problem-solving exercises. Students can easily use ChatGPT to generate solutions or code snippets. While this may save time, it can lead to surface-level understanding if students do not critically engage with the generated output. Recommendation: Design homework problems that require personalization (e.g., datasets unique to each student, reflection on learning process) and include a short self-explanation section to encourage deeper understanding. Randomized parameters or open-ended design questions also reduce direct reliance on AI.

## 3.2 Take-Home Exams

Take-home exams are designed to assess a student's ability to apply course knowledge independently over a fixed period. ChatGPT can potentially solve many exam-style questions, from theoretical derivations to programming problems. Recommendation: Use application-oriented or multi-part questions that require justification of choices, critical thinking, and reflection. Include plagiarism checks and enforce clear policies about AI tool usage. Consider oral defence of submitted answers as a verification method.

### 3.3 In-Class Quizzes and Exams

In-class assessments (closed-book) remain one of the most secure formats in preserving academic integrity, as students do not have access to AI tools during the session. Recommendation: Continue using this method for evaluating foundational concepts and technical competence. Design quizzes that test interpretation, tracing of algorithms, or debugging—areas that reveal true understanding.

### 3.4 Computer-Based Projects

These are major components of computing curricula and may include software development, data analysis, simulations, or embedded systems tasks. ChatGPT can help students by suggesting code structure, debugging tips, or even entire codebases.

Recommendation: Encourage use of AI tools for assistance (e.g., brainstorming, testing), but require detailed documentation, version control logs (e.g., Git), peer review sessions, or demonstration videos to verify authorship and understanding. Rubrics should include marks for design choices and iterative development, not just final output.

### 3.5 Design Projects

These are open-ended tasks where students design systems, interfaces, or solutions to real-world problems. AI tools can assist in ideation and generating initial prototypes or code structures. Recommendation: Require a design journal, rationale for technology selection, and a reflective report. Use oral presentations or peer reviews as part of the assessment to confirm comprehension and ownership.

### 3.6 Written Topic Reports

These assignments involve research, synthesis of literature, and technical writing. ChatGPT can generate well-written summaries and topic explanations, but may lack depth, originality, or proper referencing. Recommendation: Emphasize citation practices, ask for annotated bibliographies, and include critical analysis sections or commentary on the sources used. Employ AI-detection tools if permitted by institutional policy.

### 3.7 Oral Presentations

Students are required to present technical topics or project outcomes to peers or faculty. While AI tools can assist with scripting or structuring presentations, delivery and interaction remain human-centered. Recommendation: Include a Q&A session, peer evaluations, and content verification to assess understanding beyond prepared slides. Encourage students to discuss challenges and lessons learned rather than reading AI-generated scripts verbatim.

### 3.8 Peer Assessment and Group Work

Some computing programs incorporate collaborative or peer-assessed tasks. AI tools could be misused by individual group members without the knowledge of others. Recommendation: Monitor collaboration through milestone submissions, team logs, and individual contribution reports. Assign rotating roles to ensure shared responsibility.

### 3.9 Online Coding Platforms and Auto-Graders

Instructors increasingly use platforms like Hacker Rank, Code forces, or custom auto-graders. Students might use ChatGPT to submit syntactically correct solutions without genuine problem-solving. Recommendation: Combine auto-graded exercises with reflective questions or require explanation of the logic used. Limit copy-paste functionality where possible and incorporate code walkthroughs in class.

### 3.10 Tool-Specific Capabilities and Assessment Implications

AI tools differ significantly in their primary capabilities and the types of assessments they affect. Large language models such as ChatGPT are predominantly designed for natural language generation, explanation, and ideation, making them particularly influential in written assignments, literature reviews, and conceptual problem-solving tasks. In contrast, code-oriented tools such as GitHub Copilot focus on code completion, syntax generation, and software development workflows, posing greater integrity challenges in programming assignments and computer-based projects. Table 1 summarizes the primary capabilities of commonly used AI tools and their associated assessment risks in computing education.

Accordingly, assessment design and integrity controls should be aligned with the functional characteristics of the AI tools most relevant to each assessment type. For example, programming-related assessments benefit from live coding, oral explanation, or iterative modification tasks, whereas written assessments require critical review, comparison, and reflection on AI-generated content. Recent computing-education research confirms that generative AI tools affect assessment types differently, with text-oriented systems such as ChatGPT primarily influencing written and conceptual tasks, while code-oriented tools such as GitHub Copilot pose greater integrity challenges in programming and project-based assessments [23,24]. Recognizing these tool-specific differences provides a foundation for the targeted guidelines presented in Section 4.

**Table 1: AI Tool Capabilities and Associated Assessment Risks**

| AI Tool Type | Primary Capability | Affected Assessments | Primary Risk | Suggested Control |
|---|---|---|---|---|
| ChatGPT | Text generation, explanation | Essays, reports, homework | Superficial understanding | Reflection, comparison, oral defense |
| Copilot | Code completion | Programming assignments | Code outsourcing | Live coding, version control |
| Auto-complete IDEs | Syntax assistance | Labs, exercises | Over-reliance | Debug explanation |

## 4. Guidelines for various Assessments

In this section, we focus on computing colleges and propose a structured set of guidelines aimed at preserving academic integrity in the context of AI-assisted learning. Recognizing the diversity of assessment methods used in computing education, we outline strategies tailored to each assessment type that align with core educational objectives. These guidelines are designed not only to safeguard academic honesty but also to promote critical thinking, reinforce conceptual understanding, and deepen students' technical expertise. Similar concerns and recommendations have been raised in recent institutional and policy-focused studies, which emphasize the importance of assessment redesign, transparency, and verification in response to generative AI adoption in higher education [25–27].

For in-class or face-to-face assessments, student understanding can typically be evaluated directly under supervised conditions. However, for assessments completed outside the classroom, additional mechanisms are required to verify student authorship, comprehension, and ethical AI usage. Accordingly, Section 4.1 presents general guidelines applicable across assessment types, while Section 4.2 introduces assessment-specific recommendations commonly used in computing colleges.

## 4.1 General guidelines for Assessments

To maintain academic integrity while embracing the responsible use of AI tools in computing education, the following general guidelines are recommended for instructors across all assessment types:

### G 4.1.1 Require AI Usage Disclosure

If students utilize AI tools (e.g., ChatGPT) in completing any part of an assignment, they should be required to explicitly cite the tool and explain how it was used. This practice promotes transparency and honesty, reinforcing the values of academic integrity and responsible AI engagement.

### G 4.1.2 Incorporate Post-Submission Interviews

For any out-of-class (unsupervised) assessments, instructors should conduct brief follow-up interviews or oral defenses. During these sessions, students should be questioned on specific parts of their submitted work. A substantial portion of the grade should be based on the student's demonstrated understanding. This method helps verify authorship, discourages misuse of AI tools, and reinforces deeper learning.

### G 4.1.3 AI-Supported Assignments with Critical Review

In selected take-home assessments, instructors may intentionally allow or even encourage the use of AI tools. In such cases, students should be asked to critically analyze the AI-generated output, explain its logic, and submit a modified or enhanced version based on their own insights. The majority of the assessment grade should be allocated to the quality of analysis and enhancement, rather than the initial AI-generated response. This approach promotes critical thinking and reinforces the student's active role in the learning process.

### G 4.1.4 Introduce Oral Examinations

Where appropriate, consider converting some assessments into oral exams. Oral questioning provides a direct means of evaluating a student's comprehension, communication skills, and ability to apply concepts in real time. It also reduces the risk of academic dishonesty and fosters more personalized assessment. Recent studies suggest that oral examinations and interactive verification methods are particularly effective in validating student understanding in AI-assisted learning environments [22,25].

## 4.2 Specific Guidelines Related to Assessment Types

Different assessment types in computing education introduce distinct opportunities and risks when AI tools are used. While some assessments primarily evaluate conceptual understanding or communication skills, others emphasize technical implementation, problem-solving, or design reasoning. Consequently, a uniform integrity strategy is insufficient. The following guidelines are therefore tailored to specific assessment categories, aligning AI usage policies and verification mechanisms with the learning objectives and integrity risks associated with each assessment type.

### G 4.2.1 Tool-Specific Considerations for AI-Assisted Assessments

Different AI tools present distinct capabilities and integrity risks depending on the assessment context. For example, large language models such as ChatGPT are primarily suited for text generation, explanation, and ideation, making them more relevant to essay writing, literature reviews, and report preparation. In contrast, tools such as GitHub Copilot are specifically designed for code completion and software development, posing greater integrity challenges in programming assignments and computer-based projects.

Accordingly, assessment guidelines should account for tool-specific behaviors. Programming-related assessments should emphasize verification through live coding, oral explanations, or iterative modifications, while written assessments should prioritize critical review, comparison, and reflective analysis of AI-generated content. By aligning integrity mechanisms with the functional characteristics of different AI tools, instructors can more effectively preserve academic rigor while allowing responsible AI usage.

### G 4.2.1 Guidelines for Homework

Homework could contain a few questions or problems to be answered or solved by the students:
- Questions should be as specific as possible to have specific answers. Questions should not be vague so they could have different answers.
  A vague question such as *"Explain how sorting algorithms work"* may be easily answered using generic AI-generated explanations. In contrast, a specific question such as *"Compare the time and space complexity of Merge Sort and Quick Sort when applied to a dataset of $10^6$ elements under worst-case conditions"* requires contextual reasoning and deeper understanding. Designing questions with precise constraints and parameters reduces the effectiveness of generic AI-generated responses and encourages independent thinking.
- Problems should be as specific as possible with specific details and numbers.
- Problems should not have regular ideas but new ideas.
- Design problems should have specific requirements.
- After submitting the Homework, instructors should discuss the solutions with the students. Most of the grade should be devoted to this discussion.

### G 4.2.2 Guidelines for Programming Assignments:

Programming assignments are necessary in all computing colleges. These assignments include writing programs in different programming languages:
- Programming assignments should be with very specific requirements that force the students to think about the solution and develop it.
- Assignments better to be in class assignments.
- After evaluating the assignment, students are asked to make certain modifications in class and assign part of the grade to the modification.

### G 4.2.3 Guidelines for Literature Review Assignments:

Literature reviews are a foundational component in many computing and technical subjects. They offer students the opportunity to explore a focused topic or research problem, identify and analyze existing approaches, and compare techniques used in literature. To preserve academic integrity while leveraging the benefits of AI tools such as ChatGPT, the following guidelines are recommended:

- **Permitted Use with Clear Attribution** Students may be permitted to use AI tools like ChatGPT to assist in generating a preliminary draft of the literature review. However, the use of AI must be clearly disclosed in the submitted work, including how it was used and which sections were AI-assisted. This ensures transparency and accountability.
- **Mandatory Proofreading and Editing** Students must critically review and revise any AI-generated content. This includes checking for factual accuracy, relevance, coherence, and completeness. Blind reliance on AI-generated text is discouraged, as it may result in superficial or incorrect information.
- **Verification and Evaluation Report** In addition to the literature review itself, students should be required to submit a separate report evaluating the AI-generated content. This report should verify the correctness of the information, assess the comprehensiveness of the review, and reflect on any limitations or gaps in the AI output. It should also include a summary of what was added, modified, or removed by the student.
- **Assessment Focus on Critical Engagement** A substantial portion of the overall grade should be allocated to the student's evaluation report and the quality of enhancements made to the initial AI-generated draft. This emphasizes critical engagement, reinforces student understanding, and ensures that the final submission reflects independent thought and academic rigor.

### G 4.2.4 Guidelines for writing Essay

Essay assignments are commonly used in computing and interdisciplinary courses to assess students' ability to articulate ideas, explain technical concepts, or present critical reflections on specific topics or techniques. To integrate AI tools responsibly while preserving academic integrity, the following guidelines are recommended:

- **Controlled Use of AI Tools** Students may be instructed to use ChatGPT or similar AI tools to generate an initial version of the essay. This version should be submitted separately and clearly labeled as AI-generated content.
- **Independent Essay Writing Requirement** In parallel, students must write their own version of the essay independently, without direct assistance from AI tools. This encourages personal engagement with the topic, reinforces critical thinking, and ensures the development of individual writing skills.
- **Comparative Evaluation Report** Students should then produce a comparison report analyzing the AI-generated essay and their own version. The report should evaluate both versions for accuracy, depth, structure, originality, and completeness. This reflection encourages students to assess the strengths and limitations of AI-generated content and helps solidify their understanding of the topic.
- **Grading Emphasis on Critical Reflection** A significant portion of the grade should be awarded based on the student's independently written essay and the quality of the comparison report. This approach not only validates the student's effort but also promotes metacognitive learning and awareness of how to use AI tools responsibly and ethically.

### G 4.2.5 Guidelines for Computer-Based Projects

Computer-based projects often involve software development, data analysis, or implementation of algorithms. These assignments are ideal for assessing technical skills, problem-solving ability, and creativity. However, students may use AI tools like ChatGPT to generate or optimize code, which can lead to dependency or misrepresentation.

- **Encourage Responsible AI Use** Students may use AI tools for debugging, generating boilerplate code, or exploring alternative solutions, but this must be clearly documented in a project log or report.
- **Track Development Process** Require use of version control systems (e.g., Git) and periodic check-ins or demonstrations to track progress and verify authentic contribution.
- **Assessment Focus** Evaluate not just the final product but also the design rationale, development process, documentation, and reflection on what was learned. Oral demonstrations or live coding components can help validate student ownership.

### G 4.2.6 Guidelines for Design Projects

Design projects task students with creating solutions to open-ended problems—such as designing system architectures, user interfaces, or network topologies. AI tools can help generate initial ideas or recommend design structures.

- **Require a Design Rationale** Students should submit a design journal or report explaining their choices, iterations, and any AI-generated input.
- **Incorporate Peer Review** Mid-project peer evaluations or critiques can verify originality and prompt deeper understanding.
- **Assessment Emphasis** Focus grading on the innovation, feasibility, and reasoning behind the design, rather than just the technical execution.

### G 4.2.7 Guidelines for Written Topic Reports

Topic reports require students to synthesize information from literature or real-world technologies, often involving technical writing and analysis. AI tools can assist in drafting but may not always provide in-depth or reliable content.

- **Cite All AI Assistance** Students should document any use of AI-generated summaries or explanations and provide proper citations.
- **Emphasize Original Analysis** Require students to include personal insights, critique of sources, and a summary of key learnings. Annotated bibliographies or reflection sections can help assess understanding.
- **Plagiarism Checks** Use institutional tools to detect AI-generated content and emphasize proper paraphrasing and academic writing standards.

### G 4.2.8 Guidelines for Oral Presentations

Oral presentations assess communication, comprehension, and the ability to explain complex ideas clearly. While AI tools may help students prepare scripts or slides, the actual delivery and Q&A reflect true understanding.

- **Allow Script Assistance with Disclosure** Students may use AI to outline or draft presentations but must disclose this in a slide or verbal note.
- **Include Interactive Components** Incorporate a Q&A session or real-time problem-solving scenario to assess spontaneous thinking and subject mastery.
- **Assessment Emphasis** Focus grading on clarity, depth of explanation, delivery skills, and the student's ability to respond meaningfully to audience questions.

### G 4.2.9 Guidelines for Peer Assessment and Group Work

Collaborative assignments encourage teamwork and shared problem-solving. AI tools might be used by individual members without informing teammates, potentially leading to uneven contributions.

- **Promote Transparency** Require teams to submit a shared log detailing each member's contributions, including any use of AI tools.
- **Use Individual Reflections** Ask each student to submit a short reflection on their personal contributions, what they learned, and how the team collaborated.
- **Incorporate Milestones** Break the project into checkpoints with deliverables reviewed by peers or instructors to track authentic involvement over time.

### G 4.2.10 Guidelines for Online Coding Platforms and Auto-Graders

These platforms offer automated feedback and grading, commonly used for algorithmic problems, coding drills, and lab exercises. Students might use AI tools to copy-paste solutions without fully understanding the code.

- **Add Reflective Components** After submission, require students to explain their solution strategy, challenges faced, and how they tested their code.
- **Use Time-Limited or Live Sessions** Include periodic live coding quizzes or timed assessments to ensure independent problem-solving skills.
- **Monitor Code Patterns** Use plagiarism detection tools and manual review to detect reused or AI-generated patterns across multiple submissions.

Figure 2 illustrates how the proposed guidelines are operationalized within a structured assessment workflow. It presents the proposed assessment workflow designed to support the ethical integration of AI tools, such as ChatGPT, in computing colleges. The process begins with the identification of assessment types and the optional use of AI tools by students, provided that their usage is transparently disclosed. Assignments are guided by clear instructions and constraints, followed by student submissions that include analysis, enhancement, or personal reflection. Instructors then verify student understanding through methods such as interviews, oral exams, or code reviews. Final grades are based not only on the submitted work but also on the demonstrated comprehension, thereby reinforcing academic integrity while leveraging the educational benefits of AI assistance.

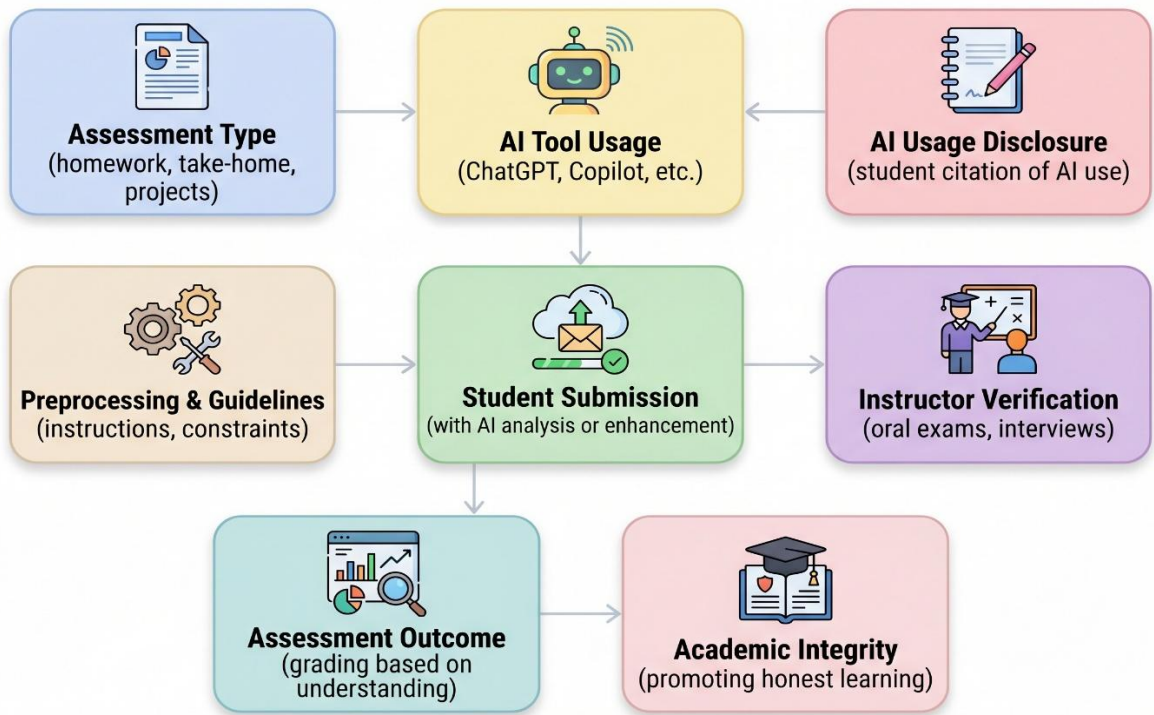

Figure 2: Proposed assessment workflow integrating AI tools in computing colleges. The diagram illustrates the process from assignment type to student submission, verification, and outcome, emphasizing responsible AI usage and academic integrity.

## 5. Formal Model of AI-Assisted Assessment Evaluation

In this section, we introduce a formal model that provides a structured mathematical framework for evaluating student assessments in the presence of AI-assisted tools. The model explicitly defines the fundamental sets, mappings, and functions involved in the assessment process, including assessment tasks, approved AI tools, and the outputs generated by both AI systems and students. By modeling AI usage, disclosure, refinement, and verification, the framework ensures transparency and traceability across all stages of assessment evaluation.

Central to the proposed model are mappings that quantify student refinement of AI-generated content, verify demonstrated understanding, and compute final grades using a weighted combination of submission quality, refinement effort, and verification outcomes. An integrity flag is introduced to ensure that only assessments satisfying both disclosure and understanding requirements are considered valid. This formalization supports fair, consistent, and academically rigorous assessment practices in educational environments where AI tools are increasingly prevalent.

### 5.1 Definitions

In this section, we define the fundamental sets and spaces involved in the assessment process as follows:
- Let $\mathcal{A}$ be the set of all assessment tasks (e.g., homework, exams, projects, etc.).

- Let $\mathcal{T}$ be the set of approved AI tools (e.g., ChatGPT, Copilot, etc.).
- Let $\mathcal{G}$ be the set of instructor-provided guidelines.
- Let $S\_AI$ and $S\_ST$ be the spaces of AI-generated drafts and student-authored submissions, respectively.
- Let $\{0,1\}$ denote Boolean outcomes (e.g., pass/fail, disclosure flag).

### 5.2 Key Mappings and Functions

Here, we introduce the core functions of the proposed model that are applied during assessment evaluation to determine grading outcomes and academic integrity through the integrity flag.

#### 5.2.1. Tool Usage and Disclosure

The tool usage, $u(a) = t$ where $a \in \mathcal{A}$ and $t \in \mathcal{T}$, indicates the student used tool $t$ on assessment $a$. The disclosure, $d(a) = \{0,1\}$, determines whether that usage is disclosed or not. In general,

$u: \mathcal{A} \to \mathcal{T} \cup \{\emptyset\}, \quad d: \mathcal{A} \to \{0,1\}$

#### 5.2.2. Preprocessing and Generation

The AI generated draft, $S\_AI = f\_AI(a, g)$ and the student's independent submission, $S\_ST = f\_ST(a)$. In general, $f\_AI: \mathcal{A} \times \mathcal{G} \to S\_AI, \quad f\_ST: \mathcal{A} \to S\_ST$.

#### 5.2.3. Comparative Analysis

This refinement factor, $r(S\_AI, S\_ST)$, measures how much the student refines the AI output. It could take a value between 0 and 1. In general, $r: S\_AI \times S\_ST \to [0, 1]$.

#### 5.2.4. Verification

The verification, $v(S_{ST}, r)$, measures student understanding in an interview or oral exam based on the student refinement. $v(S_{ST}, r) = 1$, demonstrates student understanding. In general, $v: S\_ST \times [0, 1] \to \{0, 1\}$.

#### 5.2.5. Grading Function

The grading function, $G(a)$, evaluates the grade of assessment $a$. Given that $w_1$ is weight of the score of the assessment and indicates the importance of submission quality, $w_2$ is the weight of the refinement operation importance, and $w_3$ is the weight of the verification operation importance. Then $G(a)$ can be computed as, $G(a) = w_1 \times \text{score}(S\_ST) + w_2 \times r(S\_AI, S\_ST) + w_3 \times v(S\_ST, r)$.

The weights $w_1, w_2, w_3$ are not fixed constants and may be selected by instructors or institutions based on assessment type, learning outcomes, and pedagogical priorities. For example, programming assignments may emphasize refinement and verification, whereas written reports may prioritize submission quality and demonstrated understanding. This flexibility allows the grading function to be adapted to diverse assessment contexts while preserving transparency and consistency.

#### 5.2.6. Integrity Flag

An integrity flag of assessment $a$, $I(a) = 1$ only if AI use was disclosed and the student passes verification. Otherwise, $I(a)=0$. Formally, $I(a) = d(a) \text{ AND } v(S\_ST, r)$.

Thus, we have the following result.

**Corollary 1** Let $u(a) = t$, $d(a) = \{0,1\}$ and $r(S\_AI, S\_ST) = [0,1]$. Let $v(S\_ST, r)$ demonstrates understanding of student and score($S\_ST$) reflect the student submission score, then the grade of assessment $a$, $G(a) = w_1 \times \text{score}(S\_ST) + w_2 \times r(S\_AI, S\_ST) + w_3 \times v(S\_ST, r)$, where $w_1 + w_2 + w_3 = 1$. Also, the integrity Flag $I(a) = d(a) \text{ AND } v(S\_ST, r) \in \{0,1\}$.

Corollary 1 computes the grade of an assessment and the integrity flag based on the AI submission, student submission, score, and the different weights.

### 5.2.7. Worked Example: Programming Assignment Evaluation

To illustrate the practical applicability of the proposed formal model, consider a programming assignment $a \in \mathcal{A}$ in which students are permitted to use an approved AI tool for code generation assistance, provided such usage is disclosed.

Assume a student uses an AI tool $t \in \mathcal{T}$ and properly discloses its usage, such that $d(a) = 1$. The AI-generated draft $S_{AI}$ is produced using $f_{AI}(a, g)$, while the student independently refines and submits the final solution $S_{ST}$.

The refinement factor $r(S_{AI}, S_{ST})$ is evaluated by the instructor and assigned a value of 0.6, reflecting substantial modification and improvement of the AI-generated code. During a brief oral verification or interview, the student demonstrates clear understanding of the solution design and implementation, yielding $v(S_{ST}, r) = 1$.

Let the submission quality score be $score(S_{ST}) = 0.8$, and assume weights $w_1 = 0.4$, $w_2 = 0.3$, and $w_3 = 0.3$. The final grade is computed as:
$$G(a) = 0.4 \times 0.8 + 0.3 \times 0.6 + 0.3 \times 1 = 0.82$$

Since AI usage was disclosed and verification was successful, the integrity flag is set to $I(a) = 1$. This example demonstrates how the proposed model enables transparent grading while ensuring that AI-assisted work reflects genuine student understanding and academic integrity.

### 5.3 Overall Pipeline

In this section, we give an example of steps involved in the evaluation of some assessments submitted by some students. Consider the overall pipeline for grading an assessment.

*Pipeline (Pipe_P):*
*a →(u(a), d(a)) →(f_AI(a,g), f_ST(a)) →(S_AI, S_ST) →r →v →G →I*

This sequence represents the pipeline process for an assessment *a*. Applying a tool, t, through *u* and disclosing it through *d*. This generates submissions *S_AI* and *S_ST*. These submissions need to be refined, r, then verified v. Then the grade G and integrity flag I computed.

**Theorem 2 Applying the pipeline *Pipe_P* to an assessment *a* generate the grade *G* and integrity flag *I*.**

This pipeline illustrates how AI usage, disclosure, refinement, verification, grading, and integrity evaluation are integrated into a single transparent assessment workflow.

# 6. Concluding Remarks

In light of the growing availability and widespread use of AI tools such as ChatGPT, this paper addressed the critical need to preserve academic integrity within the educational process, particularly in computing colleges. We began by outlining a set of general guidelines for instructors to adopt across various assessment formats, emphasizing transparency, critical thinking, and responsible AI usage. Building on this foundation, we then proposed specific, assessment-oriented strategies tailored to common evaluation methods in computing education. These guidelines aim to uphold the core objectives of academic instruction--namely, the development of student understanding, skills, and ethical responsibility--while adapting to the evolving technological landscape. We also introduced a formal model that provides a structured mathematical framework for evaluating student assessments in the presence of AI-assisted tools. By aligning assessment design with both educational goals and emerging AI capabilities, institutions can foster an environment that supports innovation without compromising academic standards.

## Declarations

**Ethics approval and consent to participate:**
This work complies with ethical standards. No human or animal participants were involved.

**Consent for publication:**
Not applicable.

This paper is in press for Volume 33 Issue 4 (2025) International Journal of Energy, Environment, and Economics

**Funding:**
No funds, grants, or other support was received.

**Clinical trial number:**
Not applicable.